\documentclass[aps,prl,twocolumn,floats,showpacs,superscriptaddress]{revtex4}
\usepackage{graphicx,epsfig}
\usepackage{times}
\usepackage{graphics,dcolumn,bm,float}
\usepackage{amssymb,amsmath,rotate,color}

\usepackage{graphicx}

\begin{document}
\unitlength 1 cm
\newcommand{\be}{\begin{equation}}
\newcommand{\ee}{\end{equation}}
\newcommand{\bearr}{\begin{eqnarray}}
\newcommand{\eearr}{\end{eqnarray}}
\newcommand{\nn}{\nonumber}
\newcommand{\la}{\langle}
\newcommand{\ra}{\rangle}
\newcommand{\cd}{c^\dagger}
\newcommand{\vd}{v^\dagger}
\newcommand{\ad}{a^\dagger}
\newcommand{\bd}{b^\dagger}
\newcommand{\tk}{{\tilde{k}}}
\newcommand{\tp}{{\tilde{p}}}
\newcommand{\tq}{{\tilde{q}}}
\newcommand{\eps}{\varepsilon}
\newcommand{\vk}{\vec k}
\newcommand{\vp}{\vec p}
\newcommand{\vq}{\vec q}
\newcommand{\vkp}{\vec {k'}}
\newcommand{\vpp}{\vec {p'}}
\newcommand{\vqp}{\vec {q'}}
\newcommand{\bk}{{\vec k}}
\newcommand{\bp}{{\vec p}}
\newcommand{\bq}{{\vec q}}
\newcommand{\br}{{\vec r}}
\newcommand{\bR}{{\vec R}}
\newcommand{\up}{\uparrow}
\newcommand{\down}{\downarrow}
\newcommand{\fns}{\footnotesize}
\newcommand{\ns}{\normalsize}
\newcommand{\cdag}{c^{\dagger}}

\title{Equations of motion method for triplet excitation operators in graphene}
\author{S. A. Jafari}
\affiliation{Department of Physics, Sharif University of Technology, Tehran 11155-9161, Iran}
\affiliation{School of Physics, Institute for Research in Fundamental Sciences (IPM), Tehran 19395-5531, Iran}
\author{G. Baskaran}
\affiliation{Institute of Mathematical Sciences, Chennai 600113, India}

\begin{abstract}
Particle-hole continuum in Dirac sea of graphene has a unique window underneath,
which in principle leaves a room for bound state formation in the triplet
particle hole channel [Phys. Rev. Lett. {\bf 89}, 016402 (2002)]. 
In this work, we construct appropriate triplet particle-hole operators, and using a repulsive 
Hubbard type effective interaction, we employ equations of motions to derive approximate 
eigen-value equation for such triplet operators. While the secular equation for the 
spin density fluctuations gives rise to an equation which is second order in the strength
of the short range interaction, the explicit construction of the triplet operators
obtained here shows that in terms of these operators, the second order can be factorized
to two first order equations, one of which gives rise to a solution below the
particle-hole continuum of Dirac electrons in undoped graphene. 
\end{abstract}

\maketitle

\section{Introduction}
The single particle excitations in graphene and graphite are characterized 
by a Dirac cone~\cite{Rotenberg,Bostwick,ARPESgraphite,NetoRMP}. As for the 
excitations in the two-quasi-particle sector, adding interactions may produce
bound states in, especially particle-hole channel. Such bound states exhaust
bosonic portion of excitation spectrum. In doped graphene, where an extended
Fermi surface instead of Fermi points governs the continuum of free particle-hole
excitations, the long range Coulomb forces binds initially free particle-hole pairs into spin
singlet long lived bosonic excitations known as plasmons~\cite{HwangSarma,SeyllerHREELS}.
Now, let us think of what happens in the limit where doping tends to zero?
In this limit the area of the Fermi circle becomes smaller and smaller,
so that the ratio of Coulomb energy to the kinetic energy increases,
and the single particle picture is expected to deviate from the simple
Dirac cone, whereby signatures of correlation effect are expected to 
become important in the limit of undoped graphene. 

The simplest model Hamiltonian which takes the dominant correlation
effects into account is the Hubbard model. In the light of recent
ab-initio estimates of the Hubbard $U$ in graphene, whose unscreened
value can be as large as $\sim 10$ eV~\cite{Wehling}, it is important to examine
possible consequences of such a large on-site interactions on the
physical properties of graphene. Recently extensive quantum Monte Carlo 
(QMC) study of the phase diagram of the Hubbard model on the honeycomb lattice, 
suggests spin liquid ground state~\cite{Meng} for a range 
of $U/\gamma\sim 3-4$, ($\gamma$ being the nearest neighbor
hopping amplitude). Therefore graphene is likely to be in the vicinity 
of a quantum spin liquid state~\cite{Wehling}. This scenario has been 
supported by other quantum Monte Carlo studies~\cite{Azadi}.
Our recent QMC study suggests that the collective particle-hole
excitations in $sp^2$ bonded planar systems are compatible with
a picture based on spin charge separation~\cite{Kaveh}. In this
scenario, the lowest excitations are triplet states which can 
be interpreted as two-spinon bound states. It is followed by a 
singlet excitation constructed from a doublon and a holon~\cite{Vaezi}.
Moreover, lattice gauge theory simulation of $2+1$ dimensional 
QED predicts the  critical value of the "fine structure" constant 
in graphene can be crossed in suspended graphene~\cite{Drut2009}. 
In this scenario, the ground state of graphene in vacuum is expected 
to be a Mott insulator, where in the ground state, 
the two-particle sector is dominated by long-range 
resonating valence bond correlations ~\cite{Noorbakhsh}.
Therefore, despite an intriguing simplicity of the
one-particle sector of excitations in graphene, the two-quasi-particle
sector of excitations seems to be quite involved and may have remarkable
singlet correlations in its ground state. 
Therefore it timely to revisit the nature of spin excitations in undoped
graphene~\cite{BaskaranJafari} from weak coupling side which 
is describe by a Dirac liquid fixed point~\cite{Jafari2009}.

The collective excitation considered here, will have distinct
features from plasmons, because: (i) Formation of plasmons requires
doping, while here we consider undoped graphene. (ii)
Plasmons are formed in the {\em singlet} particle-hole
channel, as a result of long range Coulomb forces. But here we assume
a short range Hubbard type interaction, and focus on the {\em triplet} channel
of particle-hole excitations. By constructing equations of motion~\cite{DemlerZhang} 
for triplet excitations formed across the valence and conduction band states
in a Dirac cone, we obtain two triplet operators whose eigen-value equations
are decoupled, and one of them displays solutions for finite values of
the short range interaction strength. We compare our derivation with a 
naive RPA-like construction of a geometric series~\cite{PeresComment}, 
and show that for the triplet operators proposed in this work, the secular 
equation decouples into two first order equations in the short range interaction 
strength, one of which always does support a solution below the particle-hole 
continuum~\cite{BaskaranJafari}. Such a decoupling can not be achieved 
for spin density fluctuation operators~\cite{PeresComment}.
Since these bosonic excitations are not precise spin density
fluctuations, their coupling to neutrons is expected to be 
less than the coupling of spin density fluctuations. 
We therefore discuss the coupling of neutrons to such excitations.

\section{Effective Hamiltonian}
As mentioned earlier, unlike plasmon (singlet) excitations, 
for which the long-range part of the Coulomb interaction is essential, 
since here we are interested in collective excitations in triplet (spin-flip) 
channel, we only need to consider the short range part of the 
interaction, as the spin-flip interactions are generated by short-range
part of the interactions. It can be shown that inclusion of longer range part of
the interactions does not lead to qualitative change in the dispersion of
spin-1 collective excitations~\cite{JafariBaskaran}.
Hence we start from the Hubbard model,
\bearr
   H&&=H_0+H_U\nn\\
   &&=-\gamma\sum_{\la i,j \ra,\sigma} \left(\ad_{i\sigma}b_{j\sigma}+
   \bd_{j\sigma}a_{i\sigma}\right)+U\sum_j n_{j\down} n_{j\up},
\eearr
where $i,j$ denote sites of a honeycomb lattice, and $\sigma$
stands for spin of electrons. 
In this model, $U\sim 10$ eV is the bare value on-site Coulomb repulsion, 
and $\gamma\sim 2.5$ eV is the hopping amplitude to nearest neighbor sites. 
To be self-contained and to fix the notations, we briefly summarize
the change of basis needed to diagonalize $H_0$.
We introduce the Fourier transforms 
\bearr
   &&\ad_{j\sigma} = \frac{1}{\sqrt N} \sum_{\bk} e^{-i\bk.\bR_j} \ad_{\bk\sigma}\nn\\
   &&\bd_{j\sigma} = \frac{1}{\sqrt N} \sum_{\bk} e^{-i\bk.\bR_{j+\delta}} \bd_{\bk\sigma}
   \nn
\eearr
where two atoms in the $j$'th unit cell are located at 
$\bR_j$ ($\in A$) and $\bR_{j+\delta}$ ($\in B$).
$N$ is the the total number of cells. The above Fourier expansion, transforms
the non-interacting part of the Hamiltonian to,
\be
   H_0=-\gamma \sum_{\la i,j\ra,\sigma} \Phi(\bk) \ad_{\bk\sigma} b_{\bk\sigma}
   +\Phi^*(\bk) \bd_{\bk\sigma}a_{\bk\sigma}
\ee
with the form factor given by,
\be
   \Phi(\bk)=e^{i\bk.\delta_1}+e^{i\bk.\delta_2}+e^{i\bk.\delta_3},
\ee
where $\delta_1,\delta_2,\delta_3$ are vectors connecting each
atom in the $A$ sub-lattice to its nearest neighbors. The phase of
the form factors are defined by 
$\gamma\Phi(\bk)=\gamma|\Phi(\bk)|e^{i\varphi_{\bk}}\equiv \eps_{\bk} e^{i\varphi_{\bk}}$,
in terms of which the hopping term becomes,
\be
   H_0= - \sum_{\bk\sigma} \eps_{\bk}
   \left(\begin{array}{cc}
   \ad_{\bk\sigma} & \bd_{\bk\sigma}
   \end{array}\right)
   \left(\begin{array}{cc}
   0			& e^{i\varphi_{\bk}}\\
   e^{-i\varphi_{\bk}}	& 0
   \end{array}\right)
   \left(\begin{array}{cc}
   a_{\bk\sigma} \\ 
   b_{\bk\sigma}
   \end{array}\right)
\ee
The following change of basis from $(a,b)$ basis to $(c,v)$ basis,
\be
   \left(\begin{array}{cc}
   a_{\bk\sigma}\\ b_{\bk\sigma}
   \end{array}\right)
   =\frac{1}{\sqrt 2}
   \left(\begin{array}{cc}
   1 			& 	1\\
   e^{-i\varphi_{\bk}}	& -e^{-i\varphi_{\bk}}
   \end{array}\right)
   \left(\begin{array}{cc}
   v_{\bk\sigma}\\ c_{\bk\sigma}
   \end{array}\right),
\ee
brings $H_0$ to diagonal format:
\be
   H_0=\sum_{\bk\sigma} \eps_{\bk} \left(\cd_{\bk\sigma}c_{\bk\sigma}-\vd_{\bk\sigma}v_{\bk\sigma}\right).
\ee
Operators $\cd_{\bk\sigma}$ and $v_{\bk\sigma}$ correspond to electron and hole
operators. For later reference we note the explicit relation connecting $\ad$ 
and $\bd$ operators to these basis is given by,
\bearr
   \ad_{j\sigma} &=& \frac{1}{\sqrt N}\sum_{\bk} e^{i\bk.\bR_j}
   \frac{1}{\sqrt 2}\left(\vd_{\bk\sigma}+\cd_{\bk\sigma} \right)\label{aj.eqn},\\
   \bd_{j\sigma} &=& \frac{1}{\sqrt N}\sum_{\bk} e^{i\bk.\bR_{j+\delta}}e^{i\varphi_{\bk}}
   \frac{1}{\sqrt 2}\left(\vd_{\bk\sigma}-\cd_{\bk\sigma} \right)\label{bj.eqn}.
\eearr

Now let us rewrite the short range Hubbard interaction in the new basis
in which $H_0$ is diagonal.
The Hubbard interaction term in the exchange channel can be written as,
\be
   H_U=-U\sum_j \ad_{j\up}a_{j\down} \ad_{j\down} a_{j\up}
   + \bd_{j\up}b_{j\down} \bd_{j\down} b_{j\up}.
\ee
Upon using Eqns.~(\ref{aj.eqn}) and (\ref{bj.eqn}) we have,
\begin{widetext}
\bearr
   \bd_{j\up} b_{j\down} &=& \frac{1}{2N}\sum_{\bk,\bk'} 
   e^{i(\bk-\bk').\bR_{j+\delta}} e^{i\varphi_{\bk}-i\varphi_{\bk'}}
   \left(\vd_{\bk\up}-\cd_{\bk\up}\right) \left(v_{\bk'\down}-c_{\bk'\down}\right)\\
   \bd_{j\down} b_{j\up} &=& \frac{1}{2N}\sum_{\bp,\bp'} 
   e^{i(\bp-\bp').\bR_{j+\delta}} e^{i\varphi_{\bp}-i\varphi_{\bp'}}
   \left(\vd_{\bp\down}-\cd_{\bp\down}\right) \left(v_{\bp'\up}-c_{\bp'\up}\right)\\
   \ad_{j\up} a_{j\down} &=& \frac{1}{2N}\sum_{\bk,\bk'} 
   e^{i(\bk-\bk').\bR_j} 
   \left(\cd_{\bk\up}+\vd_{\bk\up}\right) \left(c_{\bk'\down}+v_{\bk'\down}\right)\\
   \ad_{j\down} a_{j\up} &=& \frac{1}{2N}\sum_{\bp,\bp'} 
   e^{i(\bp-\bp').\bR_j} 
   \left(\cd_{\bp\down}+\vd_{\bp\down}\right) \left(c_{\bp'\up}+v_{\bp'\up}\right)
\eearr
Inserting the above equations in the Hubbard term, $\frac{1}{N}\sum_{j}$ produces
a momentum conservation constraint which can be satisfied by changing from
$\bk'$ and $\bp'$ to a new variable $\bq$ defined by 
\be
    \bq\equiv \bk-\bk'=\bp'-\bp,
\ee
which eventually gives
\bearr
   H_U &=& -\frac{U}{4N}\sum_{\bk\bp\bq}\left\{
   (\cd_{\bk\up}+\vd_{\bk\up})
   (c_{\bk-\bq\down}+v_{\bk-\bq\down})
   (\cd_{\bp\down}+\vd_{\bp\down})
   (c_{\bp+\bq\up}+v_{\bp+\bq\up})\right.\nn\\
   &&\left. + e^{i\varphi_\bk-i\varphi_{\bk-\bq}+i\varphi_\bp-i\varphi_{\bp+\bq}}
   (\cd_{\bk\up}-\vd_{\bk\up})
   (c_{\bk-\bq\down}-v_{\bk-\bq\down})
   (\cd_{\bp\down}-\vd_{\bp\down})
   (c_{\bp+\bq\up}-v_{\bp+\bq\up})\right\}
   \label{HU.eqn}
\eearr
\end{widetext}
Expanding the Hubbard interaction in terms of electron ($c$) and hole ($v$) operators, 
generates $32$ terms. Combining the amplitudes $1$ and
$e^{i\varphi_\bk-i\varphi_{\bk-\bq}+i\varphi_\bp-i\varphi_{\bp+\bq}}$
from first and second lines of Eq.~(\ref{HU.eqn}), leads to $16$ types of
terms with arbitrary number of electron and hole operators, whose amplitudes are 
of the form:
\be
   \gamma^{\pm}_{\bk\bp\bq} \equiv 
   \frac{1\pm e^{i\varphi_\bk-i\varphi_{\bk-\bq}+i\varphi_\bp-i\varphi_{\bp+\bq}}}{2}
\ee
As can be seen from Eq.~(\ref{HU.eqn}), for those terms containing imbalanced 
number of $c$ and $v$ operators, the amplitude of the process generated by
interaction will be $\gamma^-$, while for those where number of conduction and
valence operators are balanced, the interaction vertex will be proportional
to $\gamma^+$. Therefore in the long wave-length limit, $|\vq|\to 0$, where
$\gamma^-\to 0$, we expect the following types of terms to survive in the effective
short range interaction:
\bearr
    \tilde H_{1}=&-\frac{\tilde U}{2N}\sum_{\bk\bp\bq} 
    \gamma^+_{\bk\bp\bq}~ \cd_{\bk\up} v_{\bk-\bq\down} \vd_{\bp\down} c_{\bp+\bq\up} \\
    \tilde H_{2}=&-\frac{\tilde U}{2N}\sum_{\bk\bp\bq} 
    \gamma^+_{\bk\bp\bq}~ \vd_{\bk\up} c_{\bk-\bq\down} \cd_{\bp\down} v_{\bp+\bq\up} \\
    \tilde H_{3}=&-\frac{\tilde U}{2N}\sum_{\bk\bp\bq} 
    \gamma^+_{\bk\bp\bq}~ \cd_{\bk\up} c_{\bk-\bq\down} \vd_{\bp\down} v_{\bp+\bq\up} \\
    \tilde H_{4}=&-\frac{\tilde U}{2N}\sum_{\bk\bp\bq} 
    \gamma^+_{\bk\bp\bq}~ \vd_{\bk\up} v_{\bk-\bq\down} \cd_{\bp\down} c_{\bp+\bq\up} \\
    \tilde H_{5}=&-\frac{\tilde U}{2N}\sum_{\bk\bp\bq} 
    \gamma^+_{\bk\bp\bq}~ \cd_{\bk\up} v_{\bk-\bq\down} \cd_{\bp\down} v_{\bp+\bq\up} \\
    \tilde H_{6}=&-\frac{\tilde U}{2N}\sum_{\bk\bp\bq} 
    \gamma^+_{\bk\bp\bq}~ \vd_{\bk\up} c_{\bk-\bq\down} \vd_{\bp\down} c_{\bp+\bq\up}
\eearr
There are two more terms with their vertex strength proportional to $\gamma^+$, namely
$\cd c \cd c$ and $\vd v \vd v$ which correspond to particle-hole fluctuations solely in the
conduction or valence band, which will not contribute in the undoped graphene, as 
average occupation numbers arising from the Hartree decomposition of the equations 
of motion (see the following section) makes them irrelevant at this mean field level. 
Beyond the mean field, they are
supposed to renormalize the bare value of $U\to\tilde U$.
Therefore the tree level {\em effective} short range Hamiltonian we use in this work is,
\be
   H^{\rm eff}=H_0+\sum_{\alpha=1}^6 \tilde H_{\alpha}.
\ee
The bare value of $U\sim 4\gamma$ is expected to get renormalized to a smaller
value $\tilde U\lesssim 2.23\gamma$, beyond which an instability
occurs~\cite{JafariBaskaran}. 
The mean field factorization in the equation of motion employed here (next section) 
may lead to the underestimation of this upper value for $\tilde U$,
which is a known effect of mean field treatments~\cite{NagaosaBook}.
Therefore the physical range of parameters is limited to $\tilde U\sim 2\gamma$.
More elaborate calculations based on exact diagonalization, as well as
{\em ab-initio} quantum Monte Carlo calculation by us, supports the picture
emerging from this effective Hamiltonian~\cite{Kaveh}.
In the following section, we identify appropriate triplet operators, which 
satisfy a simple eigen-value equation, which correspond to singularities
in the one-band RPA-type susceptibility.

\section{Construction of triplet operators}
As a two-band generalization of the triplet excitation in YBCO 
superconductors~\cite{DemlerZhang}, consider two triplet operator
defined for the {\em particle-hole} channel by,
\bearr
   \cd_{\tk\up} v_{\tk+\tq\down}
   ,~~~~~~~~
   \vd_{\tk\up} c_{\tk+\tq\down}.
\eearr
These operators create triplet particle-hole excitations across the
valence and conduction bands. Therefore, by construction, these
operators are supposed to generate (triplet) excitations in undoped graphene.
To study the dynamics of these triplet excitations we calculate their 
equation of motion in a normal state. In the right hand side of terms generated by the
Hubbard interaction, we perform Hartree factorization in terms of
appropriate occupation factors and the operator under study~\cite{DemlerZhang}. 
For $\cd_{\tk\up} v_{\tk+\tq\down}$ excitations, non-zero contributions are
generated by,
\begin{widetext}
\bearr
     \left[  H_0, \cd_{\tk\up} v_{\tk+\tq\down}  \right] =& 
     \left(\eps_\tk+\eps_{\tk+\tq} \right) \cd_{\tk\up} v_{\tk+\tq\down} ,\\
     \left[ \tilde H_1, \cd_{\tk\up} v_{\tk+\tq\down}  \right] =& 
     -\frac{\tilde U}{2N} \left(\sum_\bk \gamma^+_{\bk,\tk+\tq,-\tq} \cd_{\bk\up} v_{\bk+\tq\down}\right)
     (\bar{n}^v_{\tk+\tq\down}-\bar{n}^c_{\tk\up}),\\
     \left[  \tilde H_6, \cd_{\tk\up} v_{\tk+\tq\down}  \right] =& 
     -\frac{\tilde U}{2N} \left(\sum_\bk \gamma^+_{\bk,\tk+\tq,-\tq}\vd_{\bk\up} c_{\bk+\tq\down}\right)
     (\bar{n}^v_{\tk+\tq\down}-\bar{n}^c_{\tk\up}),
\eearr
where a Hartree factorization in the right hand side has been performed to
generate the average occupation numbers~\cite{DemlerZhang}.
Similarly the non-zero contributions for triplet excitations from conduction to valence 
band after Hartree factorization become,
\bearr
     \left[  H_0, \vd_{\tk\up} c_{\tk+\tq\down}  \right] =& 
     -\left(\eps_\tk+\eps_{\tk+\tq} \right) \vd_{\tk\up} c_{\tk+\tq\down} \\
     \left[ \tilde H_2, \vd_{\tk\up} c_{\tk+\tq\down}  \right] =& 
     -\frac{\tilde U}{2N} \left(\sum_\bk \gamma^+_{\bk,\tk+\tq,-\tq} \vd_{\bk\up} c_{\bk+\tq\down}\right)
     (\bar{n}^c_{\tk+\tq\down}-\bar{n}^v_{\tk\up})\\
     \left[  \tilde H_5, \vd_{\tk\up} c_{\tk+\tq\down}  \right] =& 
     -\frac{\tilde U}{2N} \left(\sum_\bk \gamma^+_{\bk,\tk+\tq,-\tq} \cd_{\bk\up} v_{\bk+\tq\down}\right)
     (\bar{n}^c_{\tk+\tq\down}-\bar{n}^v_{\tk\up})
\eearr
The above set of results can be summarized as, 
\bearr
     \left[  H_{\rm eff}, \cd_{\tk\up} v_{\tk+\tq\down}  \right] &=& 
     \left(\eps_\tk+\eps_{\tk+\tq} \right) \cd_{\tk\up} v_{\tk+\tq\down}
     -\frac{\tilde U}{2N} (\bar{n}^v_{\tk+\tq\down}-\bar{n}^c_{\tk\up})
     \sum_\bk \gamma^+_{\bk,\tk+\tq,-\tq} 
     \left(\cd_{\bk\up} v_{\bk+\tq\down}+\vd_{\bk\up} c_{\bk+\tq\down} \right),\\
     \left[  H_{\rm eff}, \vd_{\tk\up} c_{\tk+\tq\down}  \right] &=& 
     -\left(\eps_\tk+\eps_{\tk+\tq} \right) \vd_{\tk\up} c_{\tk+\tq\down}
     -\frac{\tilde U}{2N} \left(\bar{n}^c_{\tk+\tq\down}-\bar{n}^v_{\tk\up}\right)
     \sum_\bk \gamma^+_{\bk,\tk+\tq,-\tq} 
     \left(\cd_{\bk\up} v_{\bk+\tq\down}+\vd_{\bk\up} c_{\bk+\tq\down} \right).
\eearr
Here $H_{\rm eff}=\sum_{\alpha=0}^6 H_\alpha$.
Demanding right hand side of the above equations to be $\omega_{\tq}$ times 
$\cd_{\tk\up} v_{\tk+\tq\down}$ and $\vd_{\tk\up} c_{\tk+\tq\down}$, respectively,
we obtain
\bearr
     \left(\omega_\tq-\eps_\tk-\eps_{\tk+\tq} \right) \cd_{\tk\up} v_{\tk+\tq\down} &=&
     -\frac{\tilde U}{4N} \left(\bar{n}^v_{\tk+\tq\down}-\bar{n}^c_{\tk\up}\right)
     \sum_\bk \left(1+\eta_{\bk,\tq}\eta^*_{\tk,\tq} \right)
     \left(\cd_{\bk\up} v_{\bk+\tq\down}+\vd_{\bk\up} c_{\bk+\tq\down} \right)\\
     \left(\omega_\tq+\eps_\tk+\eps_{\tk+\tq} \right) \vd_{\tk\up} c_{\tk+\tq\down} &=&
     -\frac{\tilde U}{4N} \left(\bar{n}^c_{\tk+\tq\down}-\bar{n}^v_{\tk\up}\right)
     \sum_\bk \left(1+\eta_{\bk,\tq}\eta^*_{\tk,\tq} \right) 
     \left(\cd_{\bk\up} v_{\bk+\tq\down}+\vd_{\bk\up} c_{\bk+\tq\down} \right).
\eearr
\end{widetext}
with $\eta_{\bk,\bq}\equiv e^{i\varphi_{\bk}-i\varphi_{\bk+\bq}}$, which satisfies
the property $\eta_{\bk,\bq}^{-1}=\eta^*_{\bk,\bq}=\eta_{-\bk,-\bq}$.
These set of equations suggest to define the following operators:
\bearr
   {\cal O}_{\tq}&\equiv \sum_{\bk} 
   \left(\cd_{\bk\up}v_{\bk+\tq\down}+\vd_{\bk\up}c_{\bk+\tq\down}\right),\\
   \bar {\cal O}_{\tq}&\equiv \sum_{\bk} 
   \left(\cd_{\bk\up}v_{\bk+\tq\down}+\vd_{\bk\up}c_{\bk+\tq\down}\right)\eta_{\bk,\tq}.
\eearr
The eigen-value problem for these operators become,
\bearr
   {\cal O}_\tq &= \tilde U \chi_0(\tq){\cal O}_{\tq}+\tilde U\bar\chi_0'(\tq) \bar{\cal O}_\tq,\\
   \bar{\cal O}_\tq &= \tilde U \bar\chi_0(\tq){\cal O}_{\tq}+\tilde U\chi_0(\tq) \bar{\cal O}_\tq,
\eearr
where
\bearr
   \chi_0(\tq) \!\!&=&\!\! -\frac{1}{4N}\sum_{\tk} 
   \left(\frac{\bar n^v_{\tk+\tq\down}-\bar n^c_{\tk\up}}
   {\omega_\tq-\eps_\tk-\eps_{\tk+\tq}}
   +\frac{\bar n^c_{\tk+\tq\down}-\bar n^v_{\tk\up}}
   {\omega_\tq+\eps_\tk+\eps_{\tk+\tq}}\right)\\
   \bar\chi_0(\tq) \!\!&=&\!\! -\frac{1}{4N}\sum_{\tk} 
   \left(\frac{\bar n^v_{\tk+\tq\down}-\bar n^c_{\tk\up}}
   {\omega_\tq-\eps_\tk-\eps_{\tk+\tq}}
   +\frac{\bar n^c_{\tk+\tq\down}-\bar n^v_{\tk\up}}
   {\omega_\tq+\eps_\tk+\eps_{\tk+\tq}}\right)\eta_{\tk,\tq},\nn\\
   \bar\chi_0'(\tq) \!\!&=&\!\! -\frac{1}{4N}\sum_{\tk} 
   \left(\frac{\bar n^v_{\tk+\tq\down}-\bar n^c_{\tk\up}}
   {\omega_\tq-\eps_\tk-\eps_{\tk+\tq}}
   +\frac{\bar n^c_{\tk+\tq\down}-\bar n^v_{\tk\up}}
   {\omega_\tq+\eps_\tk+\eps_{\tk+\tq}}\right)\eta^*_{\tk,\tq}.\nn
\eearr
In the last equation, since $\eta^*_{\bk,\tq}=\eta_{-\bk,-\tq}$, and the 
fractions under the parenthesis are invariant with respect to inversion
of the vectors $\bk$ and $\bk+\tq$, we conclude that $\bar\chi_0=\bar\chi_0'$.
Hence eigenvalue equations decouple into the following equations for
the symmetric and antisymmetric modes constructed from ${\cal O}$ and $\bar{\cal O}$
triplet operators:
\be
   {\cal T}_{\tq}^\pm \left( 1-\tilde U (\chi_0\pm\bar\chi_0)\right) =0.
   \label{Top.eqn}
\ee
where the normalized form of our triplet operator 
${\cal T}_\tq^\pm = {\cal O}_\tq\pm \bar{\cal O}_\tq$ is given by,
\be
   {\cal T}_{\tq}^\pm \equiv \frac{1}{{\cal N}^\pm_\tq} \sum_{\bk} 
   \left(\cd_{\bk\up}v_{\bk+\tq\down}+\vd_{\bk\up}c_{\bk+\tq\down}\right)(1\pm\eta_{\bk,\tq}).
\ee
The normalization factor satisfies,
\be
   \left({\cal N}^\pm_\tq\right)^2 = 
   4\sum_\bk\left(1\pm\cos(\varphi_\bk-\varphi_{\bk+\tq})\right)
   \label{norm.eqn}
\ee
Since we are dealing with undoped graphene, and assuming the 
temperature to be zero, the occupation numbers in the conduction
and valence bands will be $0$ and $1$, respectively. Therefore, 
the susceptibility corresponding to the above operators reduces to,
\bearr
   &&\chi_0(\tq) \pm\bar\chi_0(\tq) = \frac{1}{4N}\sum_{\bk} 
   \label{chi0boson.eqn}\\
   &&\left(\frac{1}{\omega_\tq-\eps_\bk-\eps_{\bk+\tq}+i0}
   -\frac{1}{\omega_\tq+\eps_\bk+\eps_{\bk+\tq}-i0}\right)(1\pm\eta_{\bk,\bq})\nn
\eearr

\begin{figure}[tb]
\begin{center}
\includegraphics[width=7cm,angle=0]{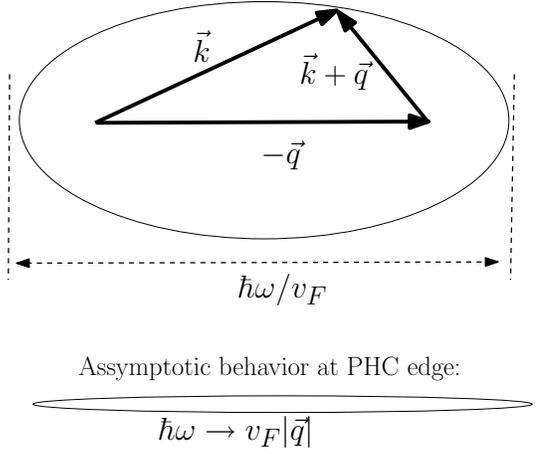}
\caption{ Points in $\vec k$-space located on the ellipse contribute
in the delta function integration of $\Im\tilde\chi_0$. In $\hbar\omega\to v_F|\vec q|\to 0$
limit the ellipse degenerates into two line segments where $\vec k$ and $\vec k'=\vec k+\vec q$
point in the opposite direction; $\theta_{\vec k}-\theta_{\vec k'}\to \pi$. Hence
the overlap factors in this limit tend to a trivial factor of $1$ and become irrelevant.
}
\label{ellipse.fig}
\end{center}
\end{figure}

In the low-energy limit where the Dirac cone linearization of the spectrum
is valid, these integrals in the the particle-hole fluctuation can be analytically 
performed~\cite{Wunsch,HwangSarma}. Otherwise they can be computed with standard
numerical procedures. Let us present here a geometric arguments based on
a very peculiar constraint in the k-space which arises from the conic spectrum.
Chiral nature of one particle eigen states, implies that the back-scattering
will not be allowed for scattering of two electrons or two holes. However,
for the scattering of a particle and a hole, the very same chiral 
nature according to which the matrix elements between a hole and
an electron state is proportional to $1-e^{i\varphi_{\bk}-i\varphi_{\bk+\bq}}$,
enhances the back-scattering between the particle and a hole.
In the small $\bq$ limit as in Fig.~\ref{ellipse.fig},
the contributions to the imaginary part of integrals in Eq.~(\ref{chi0boson.eqn})
comes from a set of points on the ellipse defined by 
$\omega_\tq=v_F(|\bk|+|\bk+\bq|)$.
Enhanced back-scattering in the particle-hole channel in the above
geometry corresponds to the limit where ellipse degenerates into 
two almost parallel line segments, i.e. the limit $\omega_\bq\approx v_F|\bq|$. 
This limit corresponds to the lower edge of the particle-hole continuum
where indeed a dominant inverse square root behavior 
(see Eq.~\ref{polar-undoped.eqn}) describes the 
non-interacting susceptibility~\cite{HwangSarma,BaskaranJafari}. Such
a geometrical constraint in the phase space could be considered as a novel
rout to 2D bosonization scheme which may find applications in systems 
with Dirac cone, such as graphene and topological insulators~\cite{HasanKane}.

Now let us study the behavior of susceptibilities corresponding to 
the two triplet operators obtained here with the aid of the
geometrical argument introduced above.
In this limit the ellipse tends to two line which enforces 
$\bk$ and $\bk+\bq$ to be in opposite directions, such that, 
$1\pm\eta_{\bk,\bq}=1\pm e^{i\varphi_{\bk}-i\varphi_{\bk+\bq}}\to 1\pm e^{i\pi}$.
Therefore in the small $\tq$ limit, $\chi_0(\tq)+\bar\chi_0(\tq)$ vanishes, and
a very large value of $\tilde U$ will be required to excite the ${\cal T}_\tq^+$. 
Hence, as long as we are interested in solutions for finite values of $\tilde U$,
we are left with the triplet operator ${\cal T}^-_\tq$ whose 
eigenvalue equation can be simplified to:
\bearr
   &&1-\tilde U \tilde \chi_0(\tq)=0,~~~~~~~~~~\tilde\chi_0(\tq)=\frac{1}{2N}\sum_\bk\label{rpa.eqn}\\
   &&\left(\frac{1}{\omega_\tq-\eps_\bk-\eps_{\bk+\tq}}
   -\frac{1}{\omega_\tq+\eps_\bk+\eps_{\bk+\tq}}\right)
   \frac{1-\cos(\varphi_\bk-\varphi_{\bk+\tq})}{2}\nn
\eearr
where we have used the symmetry of the bosonic propagator under the inversion
symmetry (in $\bk$ space) to project out the even part of the $1-e^{i\varphi_\bk-i\varphi_{\bk+\tq}}$
factor.
As we mentioned, closed form formula for the above particle-hole fluctuation can be
obtained~\cite{Wunsch, HwangSarma},
\begin{equation}
   \tilde\chi_0(\tq, \omega) =\frac{\tq^{2}}{16}
   \frac{1}{\sqrt{v_F^2 \tq^2- \omega^2}},
   \label{polar-undoped.eqn}
\end{equation}
which despite ignoring phase factors $(1-\eta_{\bk,\tq})$ 
is identical to the result obtained in Ref.~\cite{BaskaranJafari}.
The fact that ignoring these phase factors in Ref.~\cite{BaskaranJafari}
in undoped graphene, does not change the low-energy behavior 
lends on the particular geometry arising from chiral nature of 
single-particle states (Fig.~\ref{ellipse.fig}). However, for the case of doped graphene,
one has to properly take them into account, even for short range
interactions~\cite{MoradDoped}.

The eigenvalue equation for ${\cal T}_{\tq}^-$ is equivalent to divergence in the
triplet susceptibility at random phase approximation~\cite{BickersScalapino},
\be
   \chi^{\rm RPA}_{\rm triplet}(\tq,\omega)=
   \frac{\tilde\chi_0(\tq,\omega)}{1-\tilde U\tilde\chi_0(\tq,\omega)},
   \label{rpatriplet.eqn}
\ee
where retarded bare susceptibility $\chi_{0,\alpha\alpha'}(\tq,\omega)$ 
in our notation is given by the standard particle-hole form~\cite{HwangSarma},
\bearr
   \frac{1}{2N} \sum_{\bk} 
   \frac{(\bar n^{\alpha'}_{\vec{k}+\tq}-\bar n^{\alpha}_{\vec{k}})
   \left(1+{\alpha\alpha'}\cos(\varphi_{\bk}-\varphi_{\bk+\tq})\right)}
   {\hbar\omega-(\alpha'\varepsilon_{{\bk+\tq}}-\alpha\varepsilon_{\bk})+i0^{+}}.
   \label{bare-polar.eqn}
\eearr
Here $\alpha,\alpha'$ take $\pm$ values corresponding to conduction 
and valence bands, respectively~\cite{HwangSarma}. To understand the 
origin of overlap factors in this expression we note that 
the matrix elements of the scattering interaction $V$ between chiral states $(\vec k,\alpha)$ and
$(\vec k',\alpha')$ of the cone-like dispersion in graphene are given
by $\langle \vec k',\alpha'|V|\vec k,\alpha\rangle=
\tilde V(\vec k-\vec k')\left(1+\alpha\alpha' e^{i\varphi_{\vec k}-i\varphi_{\vec k'}} \right)/2 $,
where $\tilde V$ is the Fourier transform
of the scattering potential. When the above phase factors are inserted into
particle-hole bubble diagrams, give rise to the overlap factor in the free
particle-hole propagator, Eq.~\eqref{bare-polar.eqn}.
Therefore, despite that the operators considered here are not exactly 
what one expects from a local spin fluctuation operator, nevertheless,
the susceptibility corresponding to them is the particle-hole bubble.
But the main difference between spin density fluctuation and our triplet
operators is that the former satisfies a second order equation in $\tilde U$
which does not have a solution~\cite{PeresComment}, while our triplet
operators satisfy two first order equations, one of which as will be 
shown below has a solution below the continuum of free particle-hole
excitations~\cite{BaskaranJafari}.

\section{Short range versus long range interactions}
\begin{figure}[tb]
\begin{center}
\includegraphics[width=7cm,angle=0]{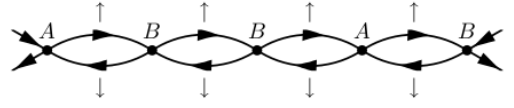}
\caption{ 
Typical geometric series for short range interactions where the
bubbles meet either on A, or B sites. Therefore internal indices
for short range interactions become identical.
}
\label{Bubbles.fig}
\end{center}
\end{figure}

To emphasize the importance of Eq.~\eqref{Top.eqn} for {\em short range 
interactions}, let us discuss how one obtains a matrix form for the 
spin density fluctuation which then leads to a second order 
secular equation~\cite{PeresComment}. 
When the range of interactions is so short that the two neighboring atoms
from two sub-lattices A, B can be resolved, an RPA like geometric series for the
susceptibility gives rise to the following equations:
\bearr
   \chi_{AA} &= \chi^0_{AA}+\tilde U\chi^0_{AA}\chi_{AA}+\tilde U\chi^0_{AB}\chi_{BA}\nn\\
   \chi_{AB} &= \chi^0_{AB}+\tilde U\chi^0_{AA}\chi_{AB}+\tilde U\chi^0_{AB}\chi_{BB}\nn\\
   \chi_{BA} &= \chi^0_{BA}+\tilde U\chi^0_{BA}\chi_{AA}+\tilde U\chi^0_{BB}\chi_{BA}\nn\\
   \chi_{BB} &= \chi^0_{BB}+\tilde U\chi^0_{BA}\chi_{AB}+\tilde U\chi^0_{BB}\chi_{BB}\nn
\eearr
where due to short range interaction $U$, the Hubbard interaction connects
$\chi^0\chi$ products in a manner that their internal indices are the same (Fig.~\ref{Bubbles.fig}).
The above set of equations decouples and gives two sets of determinants of
the following form,
\be
   \left\vert \begin{array}{cc}
      1-\tilde U\chi^0_{AA} & -\tilde U \chi^0_{AB}\\
      -\tilde U\chi^0_{BA}  & 1-\tilde U\chi^0_{BB}
   \end{array}\right\vert=0,
\ee
which is a quadratic equation obtained by constructing RPA-like
series of Feynman diagrams for particle-hole pairs propagating between 
the lattice sites. This condition corresponds to the poles in the 
RPA susceptibility of spin density fluctuations, which will not
lead to any solution (bound state)~\cite{PeresComment}. 

To see why the second order equation is peculiar to short range interactions,
consider long range interactions, for which one can construct the following 
geometric series,
\bearr
   \chi_{AA} &=& \chi^0_{AA}+W\sum_{m,m'}\chi^0_{Am}\chi_{m'A}\nn\\
   \chi_{AB} &=& \chi^0_{AB}+W\sum_{m,m'}\chi^0_{Am}\chi_{m'B}\nn\\
   \chi_{BA} &=& \chi^0_{BA}+W\sum_{m,m'}\chi^0_{Bm}\chi_{m'A}\nn\\
   \chi_{BB} &=& \chi^0_{BB}+W\sum_{m,m'}\chi^0_{Bm}\chi_{m'B}.\nn
\eearr
Here due to long-range interaction $W$, the indices $m,m'$ are not 
necessarily identical, and they can take both $A$ and $B$ values,
as the long range interaction treats both indices on the same footing.
Summing all these equations, we can see that a symmetric mode decouples
from the rest of equations, which satisfies the following equation:
\be
   1-W(\chi^0_{AA}+\chi^0_{AB}+\chi^0_{BA}+\chi^0_{BB})=0,
\ee
which is first order in the (long range) interaction strength, $W$.
Therefore, the factorization of the eigenvalue equation into first
order equations in presence of short range interactions discussed
for our triplet operators is not trivial, and is a consequence of
the peculiar form of these operators. If one writes down the 
spin density operator, e.g. Eq.~\eqref{sp.eqn}, one can see that 
the operators considered here do {\em not} precisely correspond
to spin density fluctuations. Spin density fluctuations in presence
of short range interactions give rise to a second order equation
as argued above~\cite{PeresComment}. However, a judicious choice
of ${\cal T}^\pm_{\tq}$ operators done here, manages to decouple 
an equation which otherwise is expected to be second order, into
two first order equations. 
This argument not only presents the explicit
formula for the triplet fluctuations, but also supports our
earlier prediction of neutral triplet excitations in undoped 
graphene and graphite~\cite{BaskaranJafari}.
 
\begin{figure}[tb]
\begin{center}
\vspace{-5mm}
\includegraphics[width=9cm,height=6cm,angle=0]{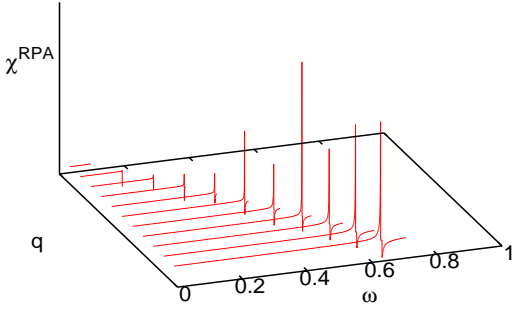}
\vspace{-8mm}
\caption{(Color online) Plot of the real part of RPA-like expression
in undoped graphene, for $\tilde U=1.9\gamma$. 
The blank region in $\omega-q$ plane corresponding to higher values
of $\omega$ is the particle-hole continuum, where the imaginary part of $\chi^0$ 
becomes non-zero. As can be seen, a singularity in the the RPA susceptibility below
the particle-hole continuum occurs, which indicates vanishing of the RPA denominator,
which is nothing but solution of the eigen-value equation for the triplet collective
excitation. Energies are in units of $\gamma$.
}
\label{undoped-overlap.fig}
\end{center}
\vspace{-4mm}
\end{figure}

 \section{Neutron scattering cross section}
 
Using Eq.~(\ref{polar-undoped.eqn}) to solve the eigen-value equation~\eqref{rpa.eqn},
gives the following dispersion for the collective triplet excitations~\cite{BaskaranJafari},
\begin{equation}
   \omega(\tq)=v_F \tq-\frac{\tilde U^2}{32v_F }\tq^3,
   \label{s1-undoped.eqn}
\end{equation}
which is valid in the limit where Dirac cone linearization applies.
When the entire band dispersion is used, one can perform the integrals
numerically. The solutions of the eigenvalue equation~\eqref{Top.eqn} for
${\cal T}^-_\tq$ operator, can be visualized as singularities in an
RPA-like susceptibility $\tilde \chi_0/(1-\tilde U \tilde \chi_0)$.
In Fig.~\ref{undoped-overlap.fig} we have plotted the real part of
the above RPA-like expression for few values of $\tq$ and a typical
value of $\tilde U=1.9\gamma$. The location of sharp divergences
in the horizontal plane in this plot represents the dispersion of
the neutral triplet collective excitation generated by ${\cal T}^-_{\tq}$.

The Dirac cone description of the electronic states of graphene equally
holds in graphite, as long as one is interested in energy scales above
the inter-layer hopping, $t_\perp\sim 50$~meV.
The Dirac cone description of the electronic states is valid in 
length scales much larger than the lattice spacing. Therefore, even
in highly oriented pyrolitic graphite (HOPG) where various planes might be
slightly rotated around the $z$-axis with respect to each other, anisotropy
in the momentum space can be safely ignored and still a Dirac cone description 
will remain valid. Hence our formulation of the spin-1 collective excitations
is not only relevant to graphene, but also will be relevant to graphite and 
HOPG at energy scales above the inter-layer hopping. 
For such bulk samples, one may think of neutron scattering to search
for the spin-1 collective excitations. 
However, since the operator corresponding to the triplet excitation
is not identical to the spin density fluctuations, the coupling of
neutrons is expected to be renormalized by appropriate
matrix elements. Therefore in this section, we consider the behavior
of neutron peak intensity in the limit of small $\tq$, with 
explicitly taking our triplet operators into account.

In polarized neutron scattering experiments, one measures,
\be
   S(\tq,\omega) = \sum_n \vert \langle 0\vert S^-_\tq \vert n\rangle \vert^2
   \delta\left(\omega-\omega_{n0} \right)
\ee
where $|0\rangle$ and $|n\rangle$ are ground and excited states of the
whole system. As discussed in this paper, a class of approximate excitations are given by
\be
   |n\rangle = \frac{1}{{\cal N}^-_\tq} {\cal T}^-_\tq |0\rangle
\ee
where the normalization factor $1/{\cal N}^-_\tq$ has been defined
by Eq.~\eqref{norm.eqn}.
Contribution of this class of excitations to the structure factor
will be given by 
\bearr
   S_{\rm triplet}(\tq,\omega) &=& \frac{1}{{\cal N}_\tq^2}
   \left|\langle 0| S^-_\tq {\cal T}^-_\tq |0\rangle \right|^2 \delta(\omega-\omega_\tq)\nn\\
   &=& \frac{1}{{\cal N}_\tq^2}
   \left|\langle 0| [ S^-_\tq, {\cal T}^-_\tq ] 
   |0\rangle \right|^2 \delta(\omega-\omega_\tq)\nn\\
\eearr
where we have used the fact that there are no triplet excitations in
the ground state: $\langle 0|{\cal T}^-_\tq=0$. 
Moreover, note that here we need the vacuum expectation value of
the ${\cal T}^-_\tq$, so that the $\vd_\up c_\down$ term gives
zero when acting on $|0\rangle$. Hence in the calculation of commutators,
we drop the $\cd_\up v_\down$ part of the triplet operator.
The spin-flip operator in the present two-band situation is given by
\bearr
   S^-_{\tq,a} &= \frac{1}{2\sqrt N} \sum_\bp
   \left(\vd_{\bp\down}+\cd_{\bp\down}\right) \left(v_{\bp-\tq\up}+c_{\bp-\tq\up}\right)\\
   S^-_{\tq,b} &= \frac{1}{2\sqrt N} \sum_\bp e^{i\varphi_\bp-i\varphi_{\bp-\tq}}
   \left(\vd_{\bp\down}-\cd_{\bp\down}\right) \left(v_{\bp-\tq\up}-c_{\bp-\tq\up}\right)
   \label{sp.eqn}
\eearr
The required commutators will become
\bearr
  \langle\left[S^-_{\tq,a},{\cal T}^-_\tq \right]\rangle &=& \frac{1}{\sqrt N} \sum_\bk
  \left(1-\eta_{\bk,\tq} \right)
  \left(\langle\vd_{\bk+\tq\down}v_{\bk+\tq\down}-\cd_{\bk\up}c_{\bk\up}\rangle\right)\nn\\
  \left[S^-_{\tq,b},{\cal T}^-_\tq \right] &=& -\left[S^-_{\tq,a},{\cal T}^-_\tq \right]^*
\eearr
At zero temperature, the conduction band is empty and the valence band 
is completely filled, so that the intensity of the mode will be
given by 
\be
   \frac{1}{{\cal N}^2_{\tq}}\left|\sum_{\bk}(1-\eta_{\bk,\tq})\right|^2
   \sim \frac{|\tq^2|^2}{\tq^2}\sim \tq^2
\ee
where the asymptotic expressions for the integrals required above
are obtained with the aid of the following expansion:
\be
   |\bk+\tq|=k+\hat k.\tq+\frac{1}{2k}[\tq^2-(\tq.\hat k)^2]+{\cal O}(\tq^3)
\ee
The $\propto \tq^2$ behavior of the neutron scattering intensity makes
the direct observation of such quanta of triplet excitations challenging
for neutron scattering experiments. Optimum spots for a neutron scattering
experiments are away from the $\Gamma$ point~\cite{JafariBaskaran}.
Moreover, due to vanishingly small binding energy of the triplet excitations
with respect to the lower boundary of the particle-hole continuum in
Fig.~\ref{undoped-overlap.fig}, the resulting neutron peak even if 
observed, maybe washed by broad spectra of the adjacent free particle-hole
pairs. Therefore a gap opening mechanism in graphite (such as proximity 
to a superconducting condensate, etc.) can be helpful
in separating the energy scales associated with the expected sharp 
resonance peak from broad features associated with the continuum of
incoherent excitations.

\section{Summary and discussions}
The secular equation obtained by Peres and coworkers~\cite{PeresComment} for
spin density fluctuations in presence of short range interactions, 
is second order in $\tilde U$, which does not
admit a solution. However, here instead of second order equation,
we obtain two set of first order equations for ${\cal T}^\pm_{\tq}$
operators, one of which (${\cal T}^+_{\tq}$) does not lead to split-off state for finite $\tilde U$,
while the other (${\cal T}^-_{\tq}$) satisfying a first order
equation leads to a dispersive triplet collective excitations whose
energy band-width is on the scale of the hopping amplitude $\gamma$. 
Such a bosonic branch of excitations might be responsible for:
(i) The life-time anomaly observed in time resolved photo-emission
spectroscopy of highly oriented pyrolytic graphite~\cite{Ebrahimkhas}.
(ii) The kink observed in the dispersion of Dirac electrons in
nearly free standing graphene samples~\cite{kink}.
(iii) The spin-flip excitations observed in artificial honeycomb 
lattice formed by quantum dots~\cite{vittore}. The interpretation of
such triplet mode as weak coupling analogue of two-spinon bound states
has been supported by some recent Monte Carlo calculations~\cite{Azadi,Kaveh}.
Despite intriguing simplicity of the Dirac cone for the single-particle
excitations of graphene/HOPG, it appears that the particle-hole sector of excitations is
likely to be more involved, and short-range and/or Heisenberg forms of
interactions maybe needed to capture the underlying singlet 
correlations~\cite{Azadi,Meng}.

\section{acknowledgments}
We thank K. Yamada for fruitful discussions, and hospitality at IMR, Tohoku University.
S.A.J. was supported by the National Elite Foundation (NEF) of Iran.

\end{document}